\documentclass[11pt]{article}

\usepackage{apacite}

\usepackage{txfonts}
\let\mathbb=\varmathbb
\DeclareSymbolFont{letters}{OML}{ztmcm}{m}{it}
\usepackage[scaled=.9]{helvet}

\usepackage[tight]{subfigure}
\usepackage{url}
\usepackage{xspace}
\usepackage{multirow}
\usepackage{booktabs}
\usepackage[margin=1in]{geometry}

\usepackage{array,graphicx}
\usepackage{float}
\usepackage{color}

\newcommand{\ie}{\emph{i.e.,}\xspace}
\newcommand{\eg}{\emph{e.g.,}\xspace}
\newcommand{\etal}{\emph{et al.}\xspace}

\newcommand{\setspace}{\setlength{\baselineskip}{15.5pt plus 1pt minus 1pt}}

\newcommand{\dsd}{SQuAD2.0\xspace}
\newcommand{\dnq}{Natural Questions\xspace}

\begin{document}

\title{Dataset vs Reality: Understanding Model Performance from the Perspective of Information Need}

\author{
	Mengying Yu\\
	School of Computer Science and Engineering,  Nanyang Technological University\\
	50 Nanyang Avenue, Singapore 639798\\
	\texttt{yume0004@e.ntu.edu.sg} \\ [10pt]
	Aixin Sun\thanks{Corresponding author.} \\
	School of Computer Science and Engineering,    Nanyang Technological University\\
	50 Nanyang Avenue, Singapore 639798\\
	\texttt{axsun@ntu.edu.sg}
}

\date{}

\setspace

\maketitle

\begin{abstract}
	\setspace
	
Deep learning technologies have brought us many models that outperform human beings on a few benchmarks. An interesting question is: \textit{can these models well solve real-world problems with similar settings (\eg identical input/output) to the benchmark datasets?}
We argue that a model is trained to answer the \textit{same information need} for which the training dataset is created. Although some datasets may share high structural similarities, \eg question-answer pairs for the question answering (QA) task and image-caption pairs for the image captioning (IC) task, they may represent different research tasks aiming for answering different information needs. To support our argument, we use the QA task and IC task as two case studies and compare their widely used benchmark datasets. From the perspective of \textit{information need} in the context of information retrieval, we show the differences in the dataset creation processes, and the differences in morphosyntactic properties between datasets. The differences in these datasets can be attributed to the different information needs of the specific research tasks. We encourage all researchers to consider the information need the perspective of a research task before utilizing a dataset to train a model. Likewise, while creating a dataset, researchers may also incorporate the information need perspective as a factor to determine the degree to which the dataset accurately reflects the research task they intend to tackle.
\end{abstract}

\setspace

\section{Introduction}
\label{sec:intro}

In the very first chapter of the Information Retrieval (IR) book, Manning \etal distinguish \textit{\textbf{information need}} from \textit{\textbf{query}}: ``An information need is the topic about which the user desires to know more, and a query is what the user conveys to the computer in an attempt to communicate the information need''~\cite{manning2008introduction}. 
When a query is entered into a search engine, the latter provides a list of potential documents that may be relevant to the user's search. The user then evaluates each document to determine if it contains the information he/she is looking for.\footnote{In our discussion, we refer to informational queries, and not navigational or transnational queries. For the latter types of queries, users often have clear expectations on the answers.} Crafting queries that accurately capture the information need is crucial, and this principle also holds true when it comes to creating datasets for training models, as the ultimate goal is to address practical tasks in the real world.

\begin{figure}[t]
    \centering
    \includegraphics[width = 2.8in]{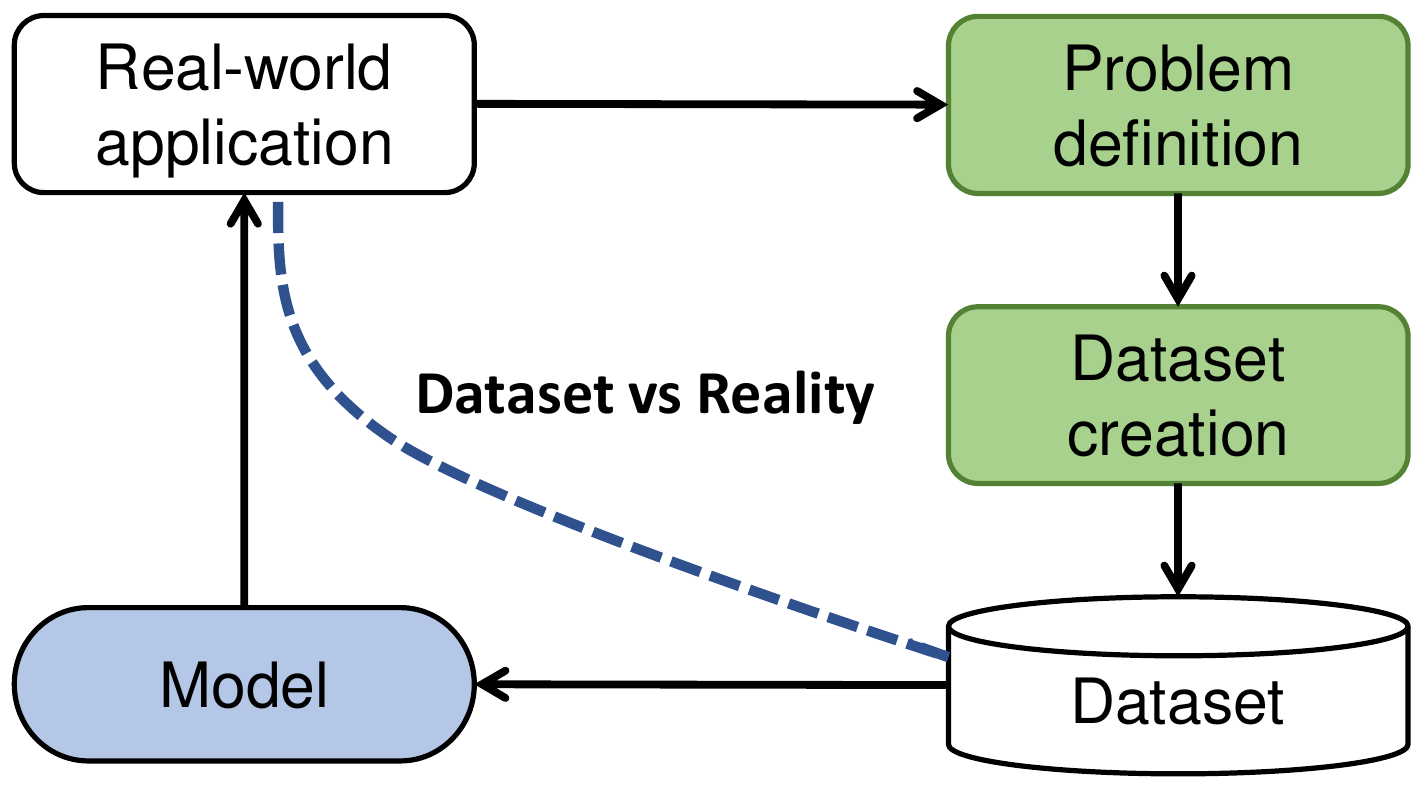}
    \caption{Overview of dataset vs reality. Model learns from dataset only and is expected to address the practical task.}
    \label{fig:overview}
\end{figure}

As illustrated in Figure~\ref{fig:overview}, before we can build a model to address a real-world problem, we need to formally formulate the problem by identifying its input/output, as well as any key constraints. To develop a model, the next essential step is to create a dataset. A dataset is used to simulate the practical task by providing inputs that are the same as, or resemble, those encountered in the real world, along with the expected outputs or ground truth labels. It is expected that the model trained on this dataset will be capable of addressing the practical problem at hand.

It is important to note that the model has no direct access to the practical problem and lacks a literal understanding of its problem definition. The model is trained to solve \textit{the problem that is ``defined'' by the dataset},  utilizing the input and corresponding expected output provided in the dataset. In this sense, a dataset is  analogous to a ``query'' in information retrieval, and the trained model plays the role of a search engine. Let us suppose that we have succeeded in training a model that is almost perfect using a dataset. Whether this trained model can effectively solve the practical problem \ie answering the ``information need'', depends largely on \textit{to what extent the dataset truly reflects the real-world problem}.

In light of the above discussion, this paper aims to conduct a systematic comparison of datasets that share similar formats (\eg input, output) and are designed for similar or disparate objectives, yet have been frequently employed as benchmarks for identical tasks. From the perspective of information need, we hope to provide a different angle in understanding a research task and also the model performance obtained on these datasets, particularly for inexperienced researchers. For instance, on some datasets, well-trained models significantly outperform human beings; on some other datasets, models are far behind human's performance. Our findings may also be helpful in guiding new dataset creation. 

As two case studies, we compare the datasets that are widely used for two research problems: question answering (QA) and image captioning (IC). Specifically, for QA, we study the differences between  SQuAD2.0 and Natural Questions (NQ) datasets. For image captioning, we compare four datasets, namely, MS-COCO, Flickr30K, SBU Captions, and Conceptual Captions. Among the six datasets, the questions in SQuAD2.0 and the image captions in MS-COCO and Flickr30K are annotated by crowdworkers; the questions in NQ and the captions in SBU Captions and Conceptual Captions are from real users or content providers.  
In our analysis, we compare the creation processes of the datasets, and try to align the dataset creation with an information need. Again, we argue that if a model is developed on a dataset, then the model is trained to answer the same information need that the dataset was created for. We perform morphosyntactic analysis on datasets with similar input/output but are created for different information needs, to show their differences. Specifically, at the word level, we compare type token ratios for lexical richness and diversity. Based on WordNet, we also compare word specificity reflected by the depth of word path in WordNet.  At the sentence level, we compare the similarity of the syntactic structure between sentences (\eg questions or image captions), computed by tree kernel similarity, and we report grammatical and/or spelling errors in sentences with an off-the-shelf language tool. 

Our morphosyntactic analysis between similar datasets shows that: (i) The texts from real users or content providers have a higher type token ratio, indicating richer and more diverse lexicon usage. Accordingly, the texts by crowdworkers are relatively easier to be handled by models, thanks to the less diverse lexicon. Based on the depth of word path in WordNet, words from real users tend to be more specific compared to the crowdsourced words, unless a dataset is purposely collected to cover such objectives during creation. (ii) In terms of sentence structure, the crowdsourced texts share similar syntactic structures, and the tree kernel similarity between these texts is much larger than that from real users or content providers.

The variations observed in the morphosyntactic analysis indicate that deep learning models may have an easier time detecting patterns in crowdsourced texts as opposed to texts written by real users.
This could be a reason why models are able to beat human performance on certain datasets, but not the other. However, as we show that the creation of these datasets may or may not share the same information need, the performance obtained on a dataset reflects the capability of deep learning models in addressing the corresponding information need that the dataset embodies. For example, the crowdsourced questions are conditioned on and are restricted to the context information that is provided during the annotation process. Therefore, these questions can only be used for testing the comprehension ability of a model or a human being. Questions of this nature do not align with the information needs typically sought by users of search engines. Conversely, captions sourced from crowdworkers provide a broad perspective of an image, making them very useful in facilitating image search through natural language.

We make two main contributions in this paper. First, we propose the perspective of information need to understand the relationship between a research task and its benchmark datasets. We argue that a model is trained to answer the same information need for which the training dataset is created.
Second, in order to support our argument, we conduct experiments on datasets created with different information needs. Specifically, we conduct morphosyntactic analysis on datasets for two research tasks QA and IC as two case studies. Through this paper, we  encourage researchers (particularly those who are new to a particular task) to look beyond the performance metric numbers obtained on some widely adopted datasets for a task, and to interpret the numbers through the lens of information need. When creating a dataset, researchers may also incorporate the information need perspective as a factor to determine the degree to which the dataset accurately reflects the research task they intend to tackle.
 
\section{Related Work}
\label{sec:related}

The rapid development of deep learning technologies has led to significant performance improvement on various tasks from vision to language. On a few benchmarks, deep learning solutions outperform human beings. Examples include  SQuAD2.0 for the reading comprehension\footnote{SQuAD2.0 leaderboard \url{https://rajpurkar.github.io/SQuAD-explorer/}} task, and MS-COCO dataset for the image captioning\footnote{MS-COCO  leaderboard: \url{https://cocodataset.org/#captions-leaderboard}} task. At the same time, researchers have also indicated that models have yet to outperform humans in the task itself~\cite{SenS20what,raji2021ai}. 

Although recent solutions bring in huge performance gains on various tasks, there are questions about the real progress that has been made in some areas~\cite{Cremonesi_Jannach_2021,raji2021ai}. The authors offer a detailed discussion on the limitations of using influential datasets to benchmark progress in machine learning~\cite{raji2021ai}. Progress made should be justified based on ``how closely our evaluations hit the mark in appropriately characterizing the actual anticipated behaviour of the system in the real world or progress on stated motivations and goals for the field''~\cite{raji2021ai}. The authors also discuss the ineffectiveness of benchmarks that measure the general ability of machine learning models in visual understanding and language understanding.  Furthermore, Miceli \etal conduct analyses from three areas: data quality, data work, and data documentation to illustrate that training models on incomplete or biased datasets may result in discriminatory outputs~\cite{miceli2022studying}. Schlangen~\citeyear{Schlangen20Targeting} answered a similar question on \textit{why models make better results on benchmark datasets constitute research progress}. In the discussion, Schlangen states that a dataset shall be verified to check ``whether the provided input/output pairs can indeed be judged correct relative to the task (in its intensional description)''. In our discussion, the perspective of information need can be considered as part of the ``intensional description''. In this paper, we offer a more concrete framework to interpret the relationships between real-world tasks, datasets, and models (see Figure~\ref{fig:overview}), through the lens of information need. We target on the concrete differences between datasets that are structurally similar but are created for answering different information needs. We argue that a model is trained to answer the information need for which the dataset was originally created for. Hence, progress made measured by benchmark datasets may or may not translate to the progress made in addressing the practical problems depending on the degree of the dataset truly reflecting the practical setting.

In our paper, we use Question Answer (QA) and Image Captioning (IC) tasks and their datasets as two case studies. For QA datasets, \cite{QAdata23Survey} offer a comprehensive survey and discussed the two different types of questions in datasets created for question answering and reading comprehension. The two types of questions are (i) information-seeking questions, and (ii) probing questions. The former is asked by users who do not know the answers, and the latter is for ``testing the knowledge of another person or machine''. In other words, the asker of a probing question knows the answer. The authors also note that the two classes of questions require different types of reasoning. The classification of question types well aligns with our perspective of information need. Nevertheless, we consider the perspective of information need is more general and can be applied to characterize datasets for many other tasks, by comparing the information need used for dataset creation and the information need for a practical task.

Regarding the image captioning case study, Torralba and Efros~\citeyear{torralba2011unbiased} argue that using captured datasets to represent the visual world and focusing solely on beating numbers on benchmarks led researchers to lose sight of their original goal. They analyzed and compared multiple datasets in computer vision, such as ImageNet, MSRC, and PASCAL, and proposed that the datasets have selection bias, capture bias, and negative set bias. These factors cause the model performance to drop significantly when training the model on one dataset and testing on another dataset. In addition, the objects in the ImageNet dataset are primarily located in the middle part of an image and are not occluded; hence the pictures do not well represent real-world images~\cite{barbu2019objectnet}. In this paper, our interest is not in image coverage or visual content. Rather, we are interested in the differences in the information need for dataset creation, reflected through the captions that come from crowdworkers and that come from content providers. The Flickr30K dataset, which is often used for image captioning, is annotated by crowdworkers, where the annotator's stereotypes and unwarranted inferences about the image can make the data biased when annotating~\cite{van2016stereotyping}. In our analysis, we do not consider the bias within annotated captions, but compare them with the captions from image uploaders. 

The differences between datasets created through crowdsourcing and datasets collected from real users or content providers, can also be considered as a kind of ``bias'' to the different information needs. Researchers in several fields have found that biases in datasets often cause errors or misunderstandings of models' capability.  In visual question answering (VQA) tasks, VQA models often suffer from language prior due to suspicious correlations between answer occurrences and certain patterns of questions~\cite{cadene2019rubi}. For instance, color of a banana is always answered as ``yellow'' regardless of the input image containing a yellow or green banana. In sequential recommendation tasks, a recommender is evaluated using datasets with genuine sequential information. However, Woolridge \etal~\cite{woolridge2021sequence} analyze the timestamp information in commonly used datasets and found that they do not represent a meaningful sequential order, especially the MovieLens dataset. As a more generic perspective, we believe information need for dataset creation and that for the practical task is a good way to evaluate the degree of a dataset truly represents the real-world scenario.

\section{Case Study 1: Question Answering Datasets}
\label{sec:QAanalysis}
We compare two question answering datasets, SQuAD2.0 and Natural Questions (NQ) from Google. Due to their structure similarity, both datasets have been widely used for evaluating question answering solutions~\cite{SenS20what}. 

\subsection{Dataset Creation}
\label{ssec:qaCreation}

\paragraph{SQuAD2.0} dataset is a large-scale question answering dataset created through crowdsourcing, on top of SQuAD1.1~\cite{rajpurkar2016squad,rajpurkar2018know}. In the original paper~\cite{rajpurkar2016squad}, the authors clearly indicate that the formulation of the SQuAD dataset is \textit{reading comprehension} (see Table~1 in~\cite{rajpurkar2016squad}). Briefly, a collection of carefully selected 536 high quality Wikipedia articles are sampled, and their paragraphs are extracted. By reading the extracted paragraphs, crowdworkers are requested to create questions based on the content, and to highlight spans in the passage as  answers. The creation of question-answer pairs was done through the Daemo platform~\cite{gaikwad2015daemo}. The crowdworkers were encouraged to post questions \textit{in their own words} and the copy-paste feature of the paragraph was disabled. SQuAD2.0 further includes 50,000 unanswerable questions, and the unanswerable question needs to be relevant to the paragraph content. When marking an unanswerable question, workers can view the corresponding paragraph. 

The SQuAD2.0 is widely used to train and evaluate question-answering models. For example, the Retro-Reader model~\cite{zhang2020retrospective} composed of the sketchy reading module and the intensive reading module has an $F_1$ score of 92.978. Google research uses the SQuAD2.0 dataset to verify that ALBERT achieves higher performance in natural language understanding with fewer parameters than BERT-large~\cite{lan2019albert}. At the time of writing, a good number of submissions report better performance than humans by either exact match or $F_1$ measure on its leaderboard.

\paragraph{Natural Questions} dataset is large scale training data for QA problems. Similar to \dsd, the answers to the questions are also from Wikipedia articles. One key difference is that the questions are real queries (anonymized and aggregated) issued to the Google search engine~\cite{kwiatkowski2019natural}. The queries are filtered heuristically to select natural questions. Wikipedia pages that appear in the top 5 search results are given to annotators to select a long answer, a short answer, or to mark null if answers cannot be found. The long answer refers to a bounding box containing the answer on the Wikipedia page, typically a paragraph or table. The short answer is a span or a set of spans, typically entities. Because the questions are real queries, the annotator also needs to judge whether a given question is good or bad; a bad question is incomprehensible and does not express the message clearly. 

Long and short answer tasks are evaluated separately. Among the state-of-the-art models, Poolingformer achieves excellent results in both tasks, with $F_1$ scores of 0.798 and 0.616 for long and short tasks, respectively~\cite{zhang2021poolingformer}. The Reflection Net~\cite{wang2020no} achieves the top ranking for the short answer task at the time of writing.\footnote{QA Leaderboard \url{https://ai.google.com/research/NaturalQuestions} As our focus in this paper is not the technical details of specific models, we refer readers to their original papers for technical details.}

\begin{figure}[t]
    \centering
    \includegraphics[width = 4in]{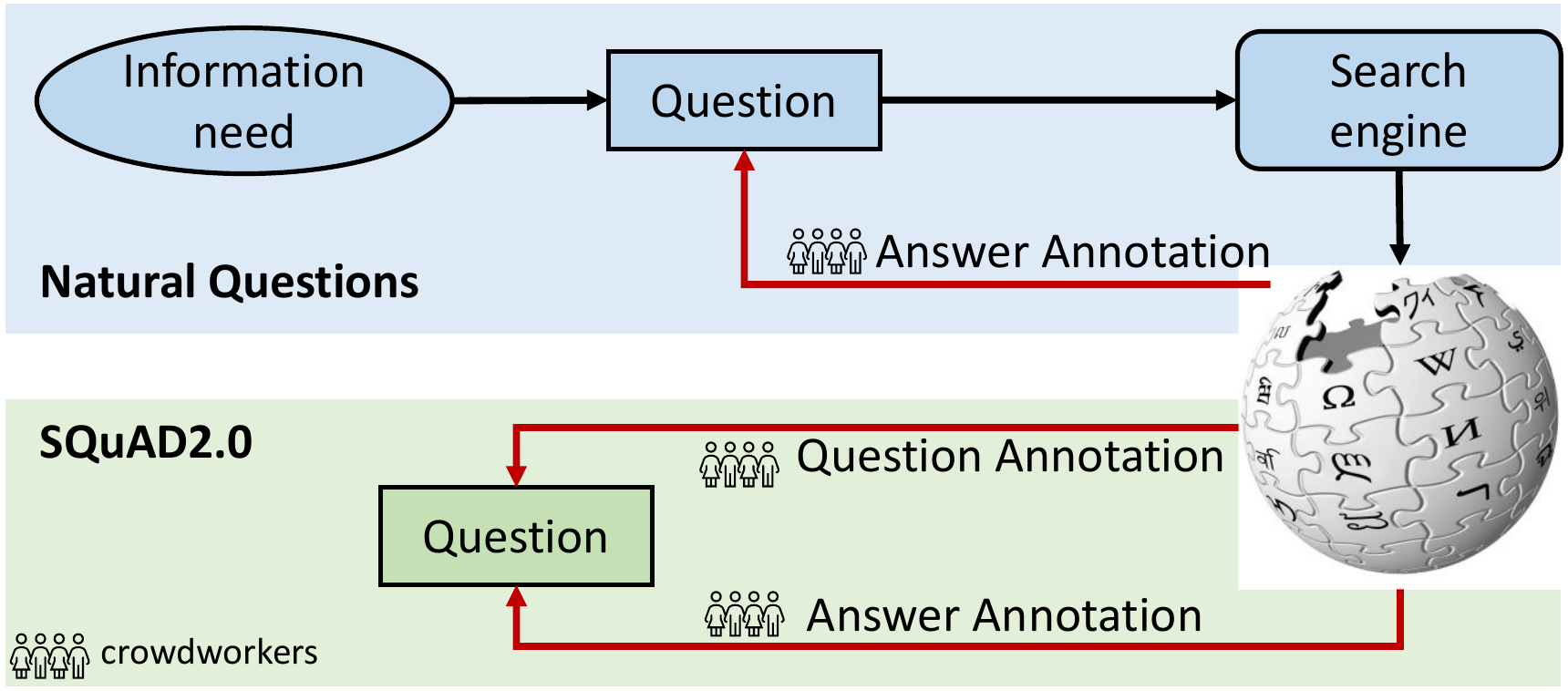}
    \caption{Information needs and data annotation. In NQ, user asks a question to learn more about the topic. In SQuAD dataset, crowdworkers ask questions to test others' (or machine's) understanding of the given context, \eg a Wikipedia article.}
    \label{fig:dsCreation}
\end{figure}

\subsection{Information Need} 

Despite the structural similarity between the two datasets, \eg both contain question-answer pairs, and all their answers are from Wikipedia pages, we see very different performances reported on leaderboards. We now compare their sources of questions, from the perspective of information need.   

\begin{table*}[th]
    \centering
     \caption{Example questions from SQuAD2.0 and Natrual Questions (NQ) with the same (short) answers. Example questions from SQuAD are restricted to the context for evaluating comprehension. For not knowning the answer, NQ questions use more generic terms like ``when'' instead of ``what season''.  }
    \label{tab:exampleQuestion}
    \begin{tabular}{l|p{5.65in}}
    \toprule
    \textbf{Dataset}& \textbf{Example questions with the answer ``football''}  \\ \midrule
    SQuAD & 1. What sports activity is featured in \underline{The Times on Mondays}?\\
& 2. What did Nigeria win a Summer Olympics gold medal for?\\
& 3. What is London's most popular athletic sport?\\ 
 NQ& 1. \underline{where} does the term muffed punt come from\\
&2. what is the most famous sport in russia\\
&3. what sports did jackie robinson play besides baseball\\\midrule
 \textbf{Dataset}& \textbf{Example questions with the answer ``winter''}  \\ \midrule
 SQuAD &1. \underline{Which season} is the most dry in Oklahoma?\\
& 2. \underline{What season} is characterized as short in Charleston?\\
&3. In \underline{what season} does New Delhi's air pollution worsen?\\
&4. During \underline{what season} is DST usually not observed because of the detriments of dark mornings?\\
& 5. The Piedmont is colder than the coast in \underline{what season}? \\
NQ
&1. when do purple martins migrate to south america\\
&2. if you live in the southern hemisphere \underline{what season} do you have in august
\\
&3. when is the newest episode of steven universe coming out\\
&4. when do deer lose their antlers in california\\
&5. when is there going to be a new steven universe episode\\
\bottomrule
    \end{tabular}
  
\end{table*}

Because the questions in \dnq are real queries, we can in general assume that each query represents an information need \textit{for which a searcher is willing to know more}, see Figure~\ref{fig:dsCreation}. Before issuing this query, the searcher does not know the answer. In this sense, the composition of this query is not restricted by or conditioned on the answer. For reading comprehension task, the annotator is given a piece of text, and based on which to ask questions and to annotate answers. Both questions and answers are annotated. The information need in this case is more aligned to ``\textit{which question-answer pairs better evaluate another reader's understanding of this piece of text}''. The questions are constrained by the given text passage and/or answers in the passage. 

Table~\ref{tab:exampleQuestion} lists sample questions from both datasets that lead to the same short answer \ie ``football'' and ``winter'', respectively. The first question for football in \dsd well reflects its constrained context \ie ``The Times on Monday''. The question is not very meaningful if the context is not given. In the contrast, in \dnq, football could be an answer to a ``where'' question. For the ``winter'' answer, questions in \dsd mostly indicate ``what season'' while the questions in \dnq mostly use ``when'' as the question issuer may not know whether the answer is a season or a month or even a specific date. We argue that questions reflect information needs, and differences in information needs in the two datasets lead to differences in the questions. As the result, the models developed on these two datasets are unlikely to fit into the same real-world application scenarios.

Next, we quantify the differences between questions in the two datasets through word-level and sentence-level analysis. In our analysis, we consider all the questions in both trainning and development, for both datasets. 

\subsection{Word-Level Analysis}
\label{ssec:qaWord}

\begin{table}
\centering
  \caption{Comparison of two Question-Answering datasets at word and sentence levels. NQ has much higher type token ratio (or higher lexical diversity), and much lower similarity between sentence structures. }
  \label{tab:QA-comparison}
  \begin{tabular}{l|rr}
    \toprule
    \textbf{Dimension} &\textbf{SQuAD2.0}&\textbf{Natural Questions}\\
    \midrule
    \textbf{Dataset} & &  \\
     ~Number of questions &142,192&315,203 \\
    ~Avg question length &  11.257& 9.370\\ \midrule
    \textbf{Word} & & \\
    ~\#Tokens & 1,600,676 & 2,953,389\\
    ~\#Token types & 45,901&59,356\\
    ~Std. Type Token Ratio & 8.860 & 11.749\\
    ~Word path depth (stem) & 8.186 (8.371) & 8.371 (8.535)\\\midrule
\textbf{Sentence} & & \\ 
~Tree Kernel Similarity & 0.088$\pm$0.002 & 0.057$\pm$0.003\\
~Tree Kernel w/o tokens & 0.267$\pm$0.008 & 0.241$\pm$0.009\\
~Error types & 533&742\\
~Error per question & 0.043 & 0.079\\
  \bottomrule
\end{tabular}
\end{table}

Considering questions in both train and development sets, 
the two datasets \dsd and \dnq contain 142,192 and 315,203 questions respectively, reported in Table~\ref{tab:QA-comparison}. 

\begin{figure}[t]
    \centering
    \includegraphics[width = 3in]{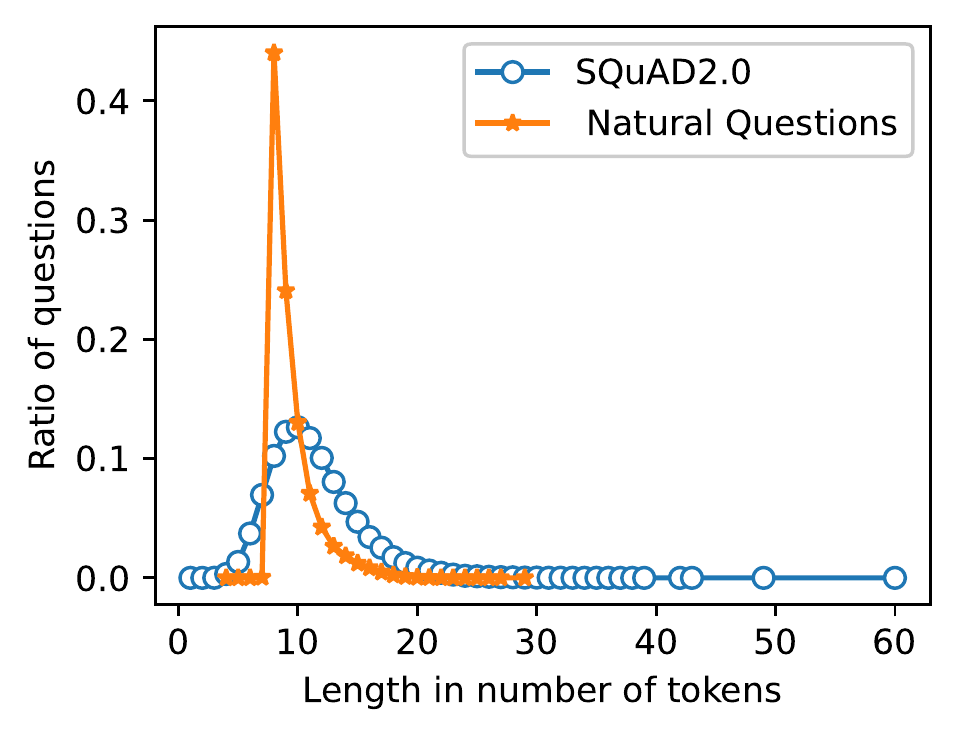}
    \caption{Question length distribution in the two datasets. Question sampling and selection process in dataset construction could be a possible reason for the distribution differences.}
    \label{fig:QA-lengthdis}
\end{figure}

The average length of the questions (in the number of tokens by SpaCy) is 11.257 and 9.370, respectively. Figure~\ref{fig:QA-lengthdis} plots the distribution of question lengths of the two datasets.  Nearly half of the questions in \dnq contain 8 words, and the distribution of question length in \dsd is relatively more smooth. Here, question sampling and selection process in dataset construction could be a possible reason for the differences in the two distributions. Hence this result is reported for reference only. Next, we compare the questions in terms of type token ratio and word specificity. 

\paragraph{Type Token Ratio.} TTR is a commonly used  measure for lexical density and richness~\cite{richards1987type}. We use the following formula to compute TTR where the number of token types is the number of unique tokens.  
$$ TTR = \frac{\#token~types}{\#tokens} \times 100$$

According to Heaps' law, there is a nonlinear relationship between token type and tokens~\cite{manning2008introduction}. Therefore, it is not appropriate to directly compare the TTR of the two datasets, due to their different sizes. Instead, we calculate the Standard Type Token Ratio - STTR~\cite{richards1987type}.  The basic idea is to divide a corpus into multiple chunks of equal size, calculate the TTR values of each chunk respectively, and then average these TTR values to obtain its STTR value. If the number of tokens contained in a chunk is fewer than the predefined number, the chunk is discarded. We use SpaCy to get tokens, and calculate the TTR for every 100,000 tokens on each dataset. 

The STTR  of the Natural Questions  is 11.749, while the value of the SQuAD2.0 dataset is 8.860 (see Table~\ref{tab:QA-comparison}). The larger STTR suggests that the lexical density and richness of real questions raised by users is much higher than that created by annotators. In general, high lexical diversity also makes a dataset more challenging.

\paragraph{Word Specificity.}  So far, our analysis of the questions shows that \dnq has a higher type token ratio, suggesting \dnq to be more diverse and challenging. Next, we use WordNet~\cite{miller1995wordnet} to measure word specificity. WordNet is a large lexical database of English where words are organized by their meanings~\cite{miller1995wordnet}. A few relations are defined in WordNet between words or more specifically synsets.  In our analysis, we use the super-subordinate relation \ie hypernymy, to measure the word specificity. Hypernymy is a relation linking more general synsets to specific synsets. Specifically, we use the \textsf{hypernym\_paths()} function in the WordNet library to calculate the average path length of nouns that appear in the questions from each dataset.  

The mean depth values of nouns with and without stemming are listed in Table~\ref{tab:QA-comparison}. Note that the average depth of nouns in the Natural Questions dataset is deeper than that in the SQuAD2.0 dataset. This result suggests that the nouns used in the questions asked by real users are more specific than those annotated by annotators. We believe this result is also consistent with the high diversity reflected by the type token ratio.

\subsection{Sentence-Level Analysis}
\label{sssec:QAsentenceDiversity}
We compare the questions between the two datasets from two perspectives: tree kernel similarity, and grammar errors.

\begin{figure}
    \centering
    \includegraphics[width =4in]{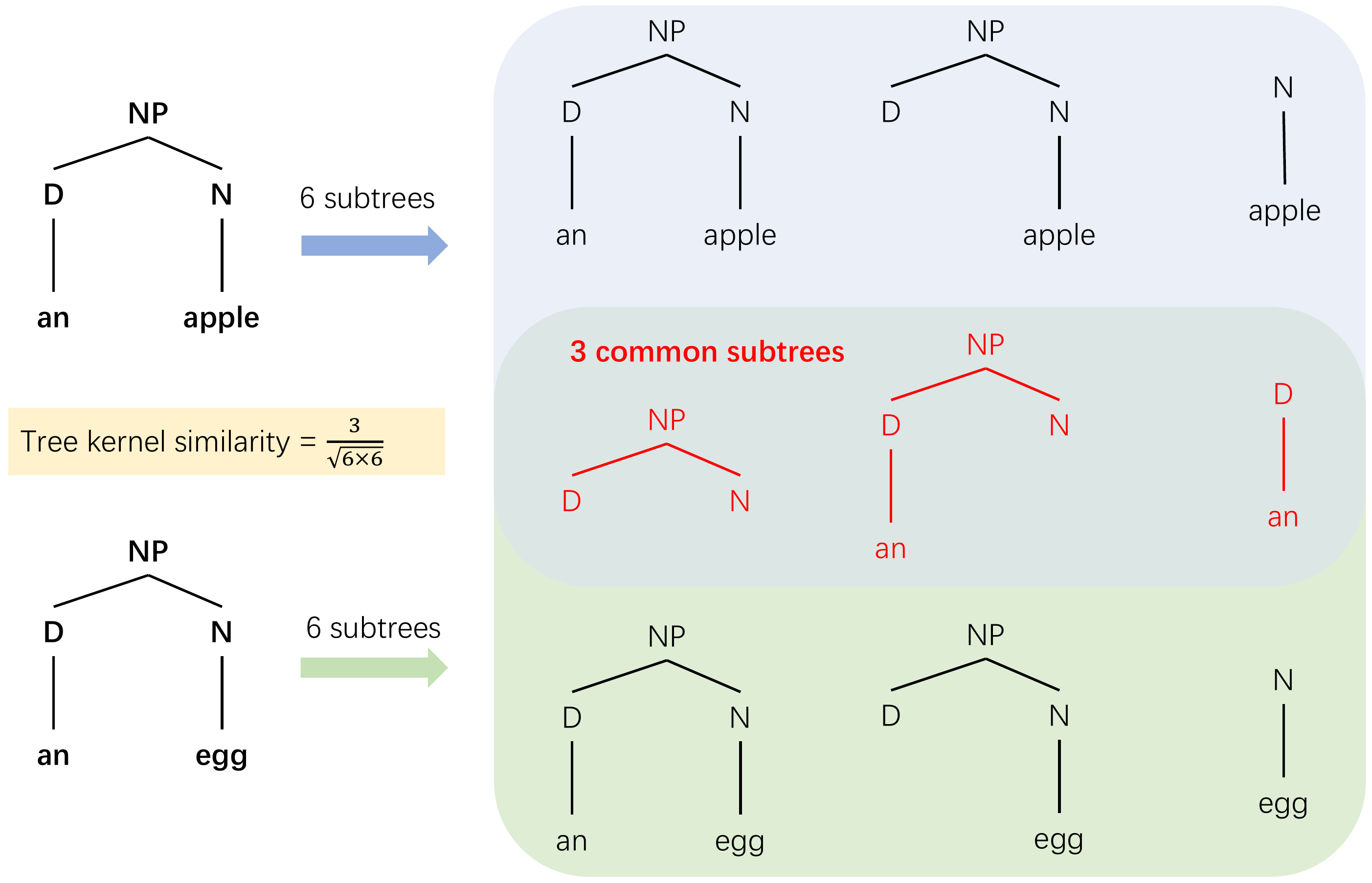}
    \caption{Illustration of tree kernel similarity between two constituency parse trees. Each parse tree has 6 subtrees and the pair has 3 common subtrees. The tree kernel similarity is the ratio between the number of common subtrees and the normalized number of subtrees in the pair, shown in the yellow box.}
    \label{fig:tk}
\end{figure}

\paragraph{Tree Kernel Similarity.}
We  use tree kernel similarity to quantify the structural similarity between questions in each dataset~\cite{collins2002new}. The central idea of tree kernel is to count the number of common subtrees between two constituency parse trees. Figure~\ref{fig:tk} gives an illustration with two simple parse trees. Each parse tree has 6 subtrees and the pair has 3 common subtrees. The tree kernel similarity is the ratio between the number of common subtrees and the normalized number of subtrees in the pair. 

We use StanfordCoreNLP to obtain constituency parse trees of questions. Then, we randomly select 10,000 parse trees and use the tree kernel to calculate their similarity. We compute two sets of values with and without the leaf-level tokens. The version with leaf-level token means that we consider the complete parse tree of a sentence. The version without leaf-level tokens is to focus solely on the sentence syntactic structure without considering the actual words, \ie the POS tags are the leaf-level nodes in a parse tree instead of the words. The average tree kernel similarity of the questions (with leaf-level token) in SQuAD2.0 is 0.088, while that of the Natural Questions dataset is 0.057. This reflects that when crowdworkers annotate questions, they are more inclined to use similar sentence patterns, leading to similar question structures in the dataset. The similarities without considering leaf-level tokens show a similar trend with 0.267 vs 0.241. In short, the questions raised by search engine users are more diverse in terms of syntax structure. 

To further compare the syntactic structure similarity between the two datasets, we conduct the unpaired t-test between the pairwise similarity values sampled from the two datasets. Specifically, we randomly select 100 questions and compute their pairwise similarity within each dataset. Then we compute the $p$-value between the two sets of similarity values and obtain $p<0.001$. We also computed the $p$-value for sample sizes of 1,000 and 10,000 questions, and in both settings $p<0.001$. This set of results indicates significant differences in the syntactic structure similarity distributions between the two datasets.

\paragraph{Grammar Accuracy.} 
Lastly, we use LanguageTool\footnote{\url{https://github.com/languagetool-org/languagetool}} to detect grammatical errors in questions from both datasets.  LanguageTool is an open-source rule-based grammar tool for comprehensive grammar and spelling error detection~\cite{naber2003rule}. Its API also serves OpenOffice's spell checker.

Because the two datasets are formatted differently, we filter out errors related to white space and upper/lowercase. As reported in Table~\ref{tab:QA-comparison}, the number of error types in Natural Questions is significantly higher than that in SQuAD2.0, as expected. Due to the different sizes of the two databases, we calculate the average number of errors per question for a fair comparison. The average number of errors per question in SQuAD2.0 is 0.043, and in Natural Questions is 0.079. In short, questions searched by users on search engines contain more grammar and/or spelling errors than questions annotated by annotators.

\begin{table*}[th]
    \centering
    \caption{Example image captions that contain keyword ``football'' from the four datasets. Crowdsourced captions provide an objective description of an image, and content providers provide additional context information that complements the image.}
    \label{tab:CaptionExample}
 \begin{tabular}{l|p{5.1in}}
    \toprule
    \textbf{Dataset} & \textbf{Captions are annotated through crowdsourcing}\\ \midrule
     \textbf{MS-COCO}
    &1. A man holding a football on a grassy field \\
    &2. A man is running while carrying a football\\
    \textbf{Flickr30K}& 1. Two football players chase after a ball.\\
    &2. A man tries to catch a football on grass surrounded by American flags.\\ \midrule
    
    \textbf{Dataset}&\textbf{Captions are from content providers}\\ \midrule
    \textbf{SBU Captions}   & 
    1. Alexander is blowing the candles on his football themed cake (made by Mom, of course!) \\
    &2. Flag football by the girls and Cheerleading by the boys. Seniors vs Juniors. The boys were deffenetly more interesting than the girls :D\\
    \textbf{C-Captions} & 1. football player celebrates scoring his side 's second goal of the game with teammates\\
    &2. rugby player of american football team during the match at american football team\\
    \bottomrule
    \end{tabular}
    
\end{table*}

\section{Case Study 2: Image Captioning Datasets}
\label{sec:IC-datasets}
For image captioning datasets analysis, we consider MS-COCO, Flickr30K, SBU Captions, and Conceptual Captions. These datasets are not created for the same purpose but have all been used for image captioning tasks.  Specifically, image captions in MS-COCO and Flickr30K are by crowdworkers, and captions in the remaining two datasets are collected together with the image, or simply image creators.  The basic statistics of the dataset regarding the number of images and captions are listed in Table~\ref{tab:IC-comparison}.

\subsection{Dataset Creation}
\label{ssec:icdatasets}

\paragraph{MS-COCO} (Microsoft Common Objects in COntext) is a large scale image dataset that can be used for image classification, object segmentation, recognition in context, and image captioning~\cite{lin2014microsoft}.\footnote{We used the 2017 version from \url{https://cocodataset.org/\#download}} Here, we are only interested in the image captions~\cite{Chen2015MicrosoftCC} which are collected through crowdsourcing. The authors of the COCO image collection carefully selected entry-level categories, \ie category labels that people often use to describe objects, and then searched images from Flickr using a combination of object categories. To label images with captions, crowdworkers were informed to describe all the important parts of the scene, and not to describe unimportant details or things that might have happened in the future or past. Each image contains about five captions. MS-COCO establishes a leaderboard for the captioning task. The Oscar (Object-Semantics Aligned Pre-training) method~\cite{li2020oscar} is currently ranked first with a BLUE-4 value of 41.7.

\paragraph{Flickr30K} dataset was created for the purpose of constructing a large scale visual denotation graph with images of everyday activities~\cite{young2014image}. Each image in this dataset has five captions from five independent crowdworkers ``who are not familiar with the specific entities and circumstances depicted''~\cite{young2014image}. Flickr30K has been used to evaluate and compare image-captioning models~\cite{you2016image}. Captions in other languages have been extended from this dataset~\cite{elliott2016multi30k,peng2016generating}.

\paragraph{SBU Captions} dataset contains 1 million images with relevant captions from Flickr~\cite{ordonez2011im2text}. Captions are from image uploaders.  During data collection, images are filtered to ensure that their text is visually descriptive. Furthermore, the text needs to contain at least one preposition for the spatial relationship of the objects in the image. The SBU Captions dataset was initially used in retrieval tasks~\cite{hodosh2013framing} and later used for image caption generation~\cite{kiros2014unifying}.

\paragraph{Conceptual Captions} dataset contains about 3.3M image and description pairs~\cite{sharma2018conceptual}. Its images and original descriptions were obtained from Web pages. Image and caption pairs were extracted, filtered, and transformed through an automatic pipeline~\cite{chambers2010flumejava}. The image descriptions are from the Alt-text elements in HTML, with careful filtering. Google AI provides a leaderboard for the image captioning task on Conceptual Captions.\footnote{Conceptual Captions \url{https://ai.google.com/research/ConceptualCaptions/}} 

\paragraph{Object Distributions.}  Images in all datasets are from the public domain; specifically, images in Conceptual Captions are from web pages and images in the other three datasets are from Flickr. Prior to performing the morphosyntactic analysis, it is imperative to take into account the distribution of objects present in the images across the four datasets. To identify the objects depicted in an image, we utilize the Faster R-CNN image object detection model~\cite{objectDetection17}. To be more specific, the Faster R-CNN model used in our experiments is a pre-trained model with the MS COCO dataset.\footnote{\url{https://github.com/open-mmlab/mmdetection/tree/master/configs/faster_rcnn}} Given the impracticality of manually verifying object detection accuracy across all datasets, we validate the detected objects through the use of ground truth captions present in each of the four datasets. That is, we consider an object is captured in an image if the object is detected by the Faster R-CNN model, and the name of the object is mentioned in the image caption. 

Subsequent to extracting the objects from the images, we sample a subset of commonly occurring objects present in all four datasets to investigate potential differences in their relative frequencies across the datasets. Specifically, we use the top 80\% most frequent objects from the Flickr30K dataset as the initial pool of sampled objects. If an object in this pool is present in all four datasets, it is included in the final set of sampled objects. Consequently, we obtain a uniform set of objects that are present in all four datasets, albeit with varying frequencies of occurrence in each individual dataset. Based on this common list, we conduct paired t-test and Table~\ref{tab:IC-Pvalues} reports the $p$-values between datasets. Observe that the $p$-values between MS-COCO, Flickr30K, and Conceptual Captions datasets are much higher than 0.05, indicating these three datasets share similar object distributions for the sampled objects.  The SBU Captions dataset on the other hand has a $p<0.001$ compared to any other dataset. The reason is that the SBU Captions dataset contains more scenery photos which resulted in a very different object distribution. It is important to note that this experiment does not provide comprehensive coverage of the range of objects present in each dataset. Instead, the results obtained from our analysis of the common pool of sampled objects indicate that three of the four datasets exhibit remarkably similar object distributions.

\begin{table}[t]
    \centering
    \caption{The $p$-values of comparing object distributions in image captioning datasets. SBU Captions has a different object distribution from the other three datasets from having many scenery photos.}
    \label{tab:IC-Pvalues}
    \begin{tabular}{l|rrrrr}
    \toprule
    p-value & MS-COCO & Flickr30K & SBU Captions & Conceptual Captions  \\\midrule
    MS-COCO & -- & 0.140 & $<0.001$ & 0.208\\ 
    Flickr30K & 0.140 & -- & $<0.001$ & 0.494\\ 
    SBU Captions & $<0.001$ & $<0.001$ & -- & $<0.001$\\ 
    Conceptual Captions & 0.208 & 0.494 & $<0.001$ & --\\
    \bottomrule      
    \end{tabular} 
\end{table}

\subsection{Information Need}

All four datasets contain image-caption pairs and all have been used to develop and evaluate image caption generation. The captions are from two different sources: through crowdsourcing in MS-COCO and Flickr30K, by content providers in SBU Captions and Conceptual Captions. Although we do not find a paper comparing the four datasets in one set of experiments, the results reported in different papers with the same measures clearly indicate that SBU Captions and Conceptual Captions are more challenging datasets compared to the other two 
datasets~\cite{li2020oscar,You_2016_CVPR,8718290,Vinyals_2015_CVPR,Changpinyo_2021_CVPR}.

Again, in this case study, we start with example captions from the datasets, and from which we explain the possible information needs in dataset creation. Table~\ref{tab:CaptionExample} lists example captions from the four datasets that contain a keyword ``football''. As expected, crowdworkers aim to objectively describe an image, similar to ``translating'' an image into a piece of text. Based on the description, if we have seen a similar scene before, we are able to visualize the picture in our minds. The \textit{information need} during the annotation process could be ``\textit{how someone else would describe this image}''. The captions generated in this way do not provide additional information on top of the visual content in a given picture, if we ignore the modality difference. In this sense, this information need is a good fit for a typical ``image search by natural language'' task. 

Captions by content providers, \eg SBU and Conceptual captions, provide \textit{additional context} to an image. Visual content and its accompanying caption complement each other and together tell the full story about the scene or event. For the same reason, captions cannot be fully discovered solely from the visual content of the image. As a result, image captioning models trained on such a dataset are less likely to give very meaningful results, if evaluated against the captions from content providers. Given that both the images and captions are sourced from content providers, there is no distinct ``information need'' linking an image to its corresponding caption in this setting, beyond the provision of supplementary information to enhance the overall presentation.

\begin{table*}
\centering
  \caption{Analysis on MS-COCO, Flickr30K, SBU Captions and Conceptual Captions (Conceptual) datasets. Datasets by content providers have higher type token ratio and lower similarity between sentence syntactic structures. }
  \label{tab:IC-comparison}
  \begin{tabular}{l|rr|rr}
    \toprule
   \textbf{Source of Caption} & \multicolumn{2}{c|}{Crowdsourced} & \multicolumn{2}{c}{Content provider} \\\midrule
    \textbf{Dimension} &\textbf{MS-COCO}&\textbf{Flickr30K}&\textbf{SBU Captions}&\textbf{Conceptual}\\
    \midrule
    \textbf{Dataset} & &  \\
    ~Number of images & 132K & 31K & 1M & 3.3M\\
     ~Number of captions & 616,767 & 158,915 & 1,000,000 & 3,334,173 \\
    ~Average caption length & 11.342 & 13.495 & 15.388 & 10.319\\ \midrule
    \textbf{Word} & & \\
    ~\#Tokens & 6,995,575 & 2,144,562 & 15,387,646 & 34,405,207\\
    ~\#Token types & 27,837 & 18,377 & 255,126 & 48,939\\
    ~Std. Type Token Ratio & 3.818 & 5.000 & 11.421 & 8.575\\
    ~Word path depth (stem) & 8.536 (8.614) & 8.451 (8.526) & 8.667 (8.791) & 8.408 (8.530)\\\midrule
\textbf{Sentence} & & \\ 
~Tree Kernel Similarity & 0.102$\pm$0.004 & 0.113$\pm$0.003 & 0.060$\pm$0.002 & 0.051$\pm$0.002\\
~Tree Kernel Similarity w/o tokens & 0.341$\pm$0.008 & 0.348$\pm$0.008 & 0.257$\pm$0.008 & 0.302$\pm$0.009\\
~Error types & 589 & 383 & 986 & 1167\\
~Error per caption & 0.077 & 0.051 & 0.069 & 0.052\\
  \bottomrule
\end{tabular}
\end{table*}

\subsection{Word-Level Analysis}
\label{ssec:ic-word}

\begin{figure}
    \centering
    \includegraphics[width =3in]{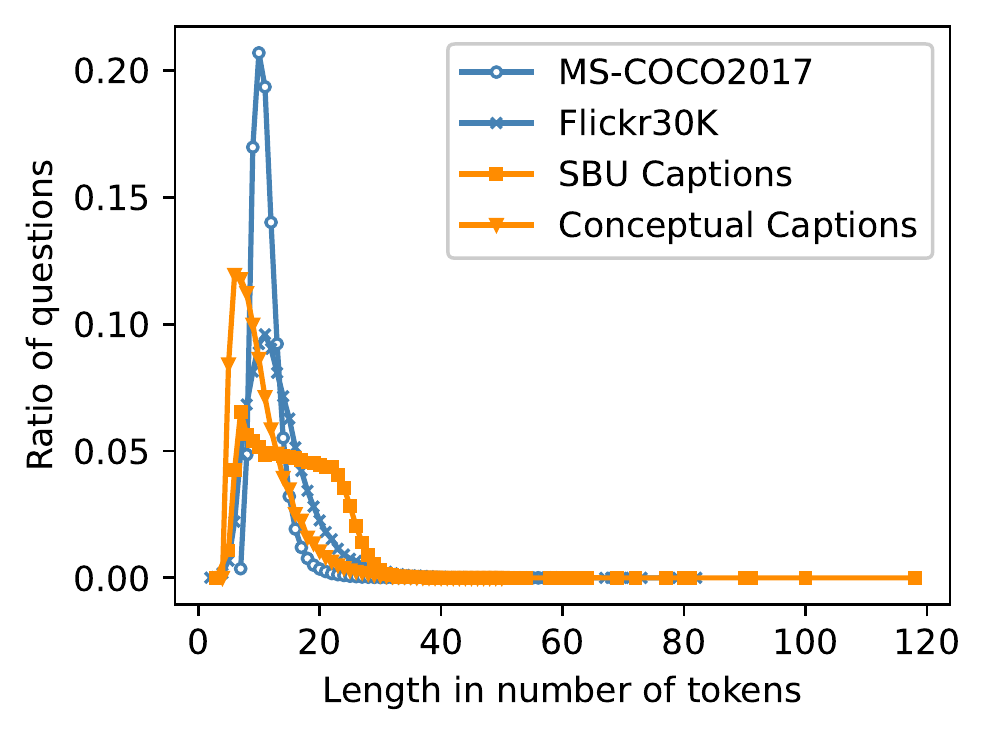}
    \caption{Caption length distribution in four datasets. The distribution is provided for reference purpose because the captions in a dataset is heavily affected by data filtering during dataset creation. }
    \label{fig:IC-lengthdis}
\end{figure}

Similarly to the QA datasets, in this section, we compare the four image captioning datasets at the word level. In terms of caption length distribution, the captions in the four datasets range from 10.3 words on average in Conceptual Captions to 15.4 words in SBU Captions. Again, because the caption lengths are heavily affected by data filtering and instructions to crowdworkers, we show the length distributions in Figure~\ref{fig:IC-lengthdis} for comprehensiveness and reference purpose only.

The standard type token ratios of MS-COCO, Flickr30K, SBU Captions, and Conceptual Captions datasets are 3.818, 5, 11.421, and 8.575 respectively, listed in Table~\ref{tab:IC-comparison}. Clearly, the captions collected from content providers contain richer and more diverse lexicons. In particular, SBU Captions has comparable STTR with Natural Questions, and both their texts are from real users. Note that, the number of captions in SBU Captions datasets is not the largest but its number of token types is significantly larger than others (see Table~\ref{tab:IC-comparison}). 

With the highest STTR, it is expected that the words in SBU Captions are more specific, with the deepest word path by WordNet. MS-COCO has a relatively deep word path as well because the images are purposely collected to cover concrete objects commonly seen in our daily life. 

\subsection{Sentence-Level Analysis}
\label{sssec:sentenceDiversity}

MS-COCO and Flickr30K share similar tree kernel similarity of 0.102 and 0.113 respectively. Both similarity values are much higher than that in SBU  and Conceptual Captions for 0.06 and 0.051 respectively. We also conduct the unpaired t-test on the pairwise tree kernel similarities computed from the randomly sampled captions in each dataset, and all $p$-values obtained are $<0.001$ between crowedsourced dataset and the dataset using user provided content. Again, tree kernel similarity measures similarity between sentence syntactic structures in the form of a parse tree. If not considering tokens in tree kernel similarity computation, the similarities from the two crowdsourced datasets remain much larger than the two datasets where data are collected from content providers. In short, captions written by content providers are richer in terms of sentence structure.

We also applied the same LanguageTool on the four image captioning datasets to detect grammar and spelling errors. However, due to differences in dataset filtering, the errors detected do not show a clear pattern in these experiments. The errors per caption are reported in Table~\ref{tab:IC-comparison} last row, where errors related to upper/lower cases and white spaces are not counted. If considering such errors, then the average number of errors per caption in SBU Captions and Conceptual Captions datasets is significantly higher.

\section{Discussion}
\label{sec:discuss}

Nowadays, leaderboards from large scale datasets have attracted significant attention from many researchers. Although improvements in leaderboard performance are indicative of progress made on a dataset, it is unclear whether these gains arise from the model's ability to address similar real-world tasks or from the model's improved fit to the peculiarities of the dataset itself. In this paper, we compare datasets through the lens of information need to which each dataset was designed to answer. We show that, even if two datasets demonstrate very similar structural similarity, and a model can be easily adapted from one dataset to another, the two datasets may represent two completely different kinds of real-world problems, because they are designed to answer two different information needs. Therefore, the trained models are meant to answer different information needs. As authors indicated in their dataset paper, SQuAD was designed for reading comprehension~\cite{rajpurkar2016squad} and Natural Questions dataset is to provide a large scale QA dataset~\cite{kwiatkowski2019natural} where the questions are asked by users who want to know more about the topics. The datasets were indeed created to answer two completely different information needs. We argue that the models trained mean to answer their corresponding information needs as well. 

Our detailed analysis of the datasets covers both word-level diversity and sentence-level diversity among them. Specifically, datasets that have been annotated by crowdworkers exhibit significantly lower standard type token ratio values compared to datasets that have been collected from real-world settings. Consequently, a trained model has a greater likelihood of fitting a crowdsourced dataset more effectively. At the sentence level, the similarity of sentence structure among annotated datasets is higher than sentences collected from real users. For image captions from content providers, they cover additional context information about the image which is not observable from the image's visual content. Again, the similar sentence structure makes crowdsourced datasets relatively easier to model. In short, the differences between datasets revealed in our analysis partially explain the high and low performances obtained on different datasets \eg from their leaderboards. Our analysis results are unsurprising given that crowdworkers are required to adhere to specific dataset creation instructions in order to maintain data quality. However, our key focus remains the information need for the dataset creation. Note that, while the dimensions employed in our analysis may have potential implications for predicting task difficulty~\cite{MishraBC13Predicting}, it is not within the scope of our present investigation to delve into the issue of task difficulty prediction. We leverage the outcomes of our analysis to substantiate our contentions regarding the distinct information needs and the degree to which a dataset can effectively capture the pertinent information needs. In the event that two datasets are not devised with the purpose of addressing identical information needs, any direct comparison of model performance would be rendered incongruous, since the models in question would have been trained to tackle divergent real-world problems.

We remark that our analysis and discussion are not to discourage dataset creation through crowdsourcing. For instance, the objective descriptions in MS-COCO and Flickr30K effectively facilitate  natural language search on images. Apart from the person who creates an image, individuals who view the image typically possess a comparable objective perception of its visual elements. In this context, the captions sourced from the crowd serve as an accurate reflection of the real-world problem scenario, as they correspond to a congruent information need. The extent to which the images contained in a dataset accurately depict those encountered in the real world is a distinct topic from the matter under discussion.

Lastly, we acknowledge that understanding the relationship between model, dataset, and a practical task is complex and complicated. In this paper, we put forth a proposition to employ the information need perspective from an Information Retrieval framework as a means to comprehend their interconnections. Nevertheless, we believe information need could be one of many possible perspectives to approach this complex topic. Moreover, question answering and image captioning are just two of the many NLP and NLU tasks. The extent to which the information need perspective can be readily utilized to elucidate other tasks remains inadequately investigated.

\section{Conclusion}
\label{sec:conclude}
Recent years have witnessed a trend in developing datasets to enable many more interesting studies on various problems. This is evidenced by the resource papers in SIGIR, and the datasets and benchmark track in NeurIPS conferences. At the same time, there is also a trend to reconsider model performance, as models outperform human benchmarks on datasets for various tasks. In this paper, we discuss dataset vs reality from the perspective of information need. We support our discussion with word-level and sentence-level analysis of datasets that are similar in format and have been used for similar tasks. We believe that all datasets are created as the result of great effort from researchers, and we are not coming to a conclusion on which dataset should or should not be used. We are also not meant to discourage dataset creation through crowdsourcing. Our aspiration is that our analysis will provide insight into how to improve the comprehension of model performance on datasets, particularly with regards to the degree of congruence between the information need of the dataset and that of a real-world application. Stated differently, we urge researchers to contemplate the specific real-world problem that a dataset genuinely embodies, in terms of the corresponding information need, before incorporating it into experiments. Similarly, we advocate for researchers to carefully assess whether the methodology employed in the creation of a dataset is capable of guaranteeing that the information need reflected in the resultant dataset genuinely aligns with that of the practical task.

\bibliographystyle{apacite}
\bibliography{ref-clean}

\end{document}